\documentclass{aa}
\usepackage[varg]{txfonts}
\usepackage{graphicx}   % Including figure files
\usepackage[colorlinks, linkcolor=blue, urlcolor=blue, citecolor=blue]{hyperref}
\usepackage{booktabs}
\usepackage{ulem}
\def\beq{\begin{equation}}
\def\enq{\end{equation}}
\def\bea{\begin{eqnarray}}
\def\ena{\end{eqnarray}}

\begin{document}

\title{Hints of a universal width-energy relation for classified fast radio bursts}

\author{Di Xiao\inst{\ref{inst1},\ref{inst3}}
\and Zi-Gao Dai \inst{\ref{inst2},\ref{inst3}}}

\institute{Purple Mountain Observatory, Chinese Academy of Sciences, Nanjing 210023, People's Republic of China \label{inst1}
\\
\email{dxiao@pmo.ac.cn}
\and
Department of Astronomy, University of Science and Technology of China, Hefei 230026, China \label{inst2}
\and
School of Astronomy and Space Science, Nanjing University, Nanjing 210023, China \label{inst3}}

\date{Received XXX / Accepted XXX}

\abstract{The total available sample of fast radio bursts (FRBs) has been  growing steadily in recent years, facilitating the study of FRBs from a statistical point of view. At the same time, the classification of FRBs is currently an imperative issue. We propose that the brightness temperature of bursts can serve as an ideal criterion for classification. In this work, we gather the available data for all localized FRBs and we find a positive relation between the intrinsic pulse width and burst energy, $T_{\rm i}\propto E_\nu^{0.25}$, for three repeating FRBs that is similar to that of our previous work using FRB 20121102A data alone. The critical line $T_{\rm B,cri}$ is found to vary for different FRBs, which may reflect the differences in source properties. This relation can put strong constraints on mainstream radiation mechanisms. It is evident that neither the coherent curvature radiation or synchrotron maser radiation have the capability to reach the high brightness temperature required to reproduce this relation.}

\keywords{Methods: statistical -- Radiation mechanisms: non-thermal}

\titlerunning{Width-Energy Relation for Classified FRBs}
\authorrunning{D. Xiao and Z. G. Dai}

\maketitle

\section{Introduction}
\label{sec1}
Fast radio bursts (FRBs) are flashing radio signals discovered fifteen years ago \citep{Lorimer2007}. Following a decade of research, the number of known FRBs is now growing rapidly thanks to the development of radio facilities and detection technology. There are hundreds of events recorded at present, as the Canadian Hydrogen Intensity Mapping Experiment (CHIME) Collaboration reported in their first FRB catalog last year \citep{CHIME2021cat}. With such a large sample, it is now possible to study FRB in a statistical context. Since this field is still at an early stage, there are many unknowns around the nature and properties of FRBs, and population study is expected play an important role in the future. Many aspects could be explored statistically, for instance, the (cumulative) distribution of observed quantities can give us some information on FRB luminosity function, event rate, or local environment \citep{Macquart2018a, Macquart2018b}. Currently, individual FRBs appear distinct in many aspects such as their pulse morphology, polarization property, and spectro-temporal behavior, however, they should have something in common if they belong to a single population. These notions point to a crucial open question of population study regarding the classification of FRBs and whether there are two or more FRB populations.

The existing classification based on repeating behavior is phenomenological. At the moment, the majority of FRBs in the catalog are one-off events. However, selection bias could play a role here, as it may make a repeater seem non-repeating. If subsequent bursts are narrower or weaker than the original one, they could possibly being missed from an observational point of view \citep{Connor2018, Palaniswamy2018}. For instance, the repetition of FRB 20171019A could not be identified until two very faint bursts were detected, which were found to be $\sim590$ times weaker than the discovery burst \citep{KumarPra2019}. This strongly implies that some repeaters may have been misidentified as non-repeaters. It is now under debate whether all FRBs do indeed exhibit repetitive behaviors \citep{Caleb2019, Lu2020a}. Several works have proposed  using a repetition fraction to study this issue, however, it will take some time to verify whether this fraction approaches unity or not with the accumulation of observing time \citep{Ai2021,Gardenier2021}. In this sense, it is not very useful to use repetition in classifying FRBs \citep[e.g.,][]{ZhangKJ2022,ZhongSQ2022} and it is necessary to find a more physical criterion as a basis \citep[e.g.,][]{GuoHY2022}. 

As we pointed out in a recent paper \citep[][hereafter Paper I]{XiaoD2022a}, brightness temperature could serve this  function since it is related to the radiation mechanism directly. Delving further into the spectral luminosity-duration phase space, different radio transients cluster in different regions characterized by their brightness temperature \citep{Nimmo2022}. We have drawn a dividing line for ``classical'' FRBs and applied this classification method to the large sample of FRB 20121102A observed by Five-hundred-meter Aperture Spherical radio Telescope \citep[FAST,][]{LiD2021}. A positive power-law relation has been found for burst width versus fluence (or energy). However, there are doubts as whether this method can be extended to a larger number of FRBs and the universality of this relation has been challenged. This work is aimed at exploring this issue and the paper is organized as follows. We introduce the statistical method and present the results in Sect. \ref{sec2}, whereby hints of a universal relation between pulse width and energy for classified FRBs are found. The implication of this relation on FRB radiation mechanism is explored in Sect. \ref{sec3}. We conclude with our discussion and conclusions in Sect. \ref{sec4}.

\section{Method of FRB classification}
\label{sec2}

The brightness temperature of an FRB is:
\bea
T_{\rm B}&=&F_\nu d_{\rm A}^2/2\pi k (\nu T)^2\nonumber\\
&=&1.1\times10^{35}{\,\rm K}\,\left(\frac{F_\nu}{\rm Jy}\right)\left(\frac{\nu}{\rm GHz}\right)^{-2}\left(\frac{T}{\rm ms}\right)^{-2}\left(\frac{d_{\rm A}}{\rm Gpc}\right)^2,\nonumber\\
\label{eq:T_B}
\ena
where $F_\nu$ is the flux density, $\nu$ is the emission frequency, $T$ is the pulse width, and $d_{\rm A}$ is the angular diameter distance \citep{Zhang2020c,XiaoD2021}. Distance information is a prerequisite for obtaining $T_{\rm B}$, therefore we adopted all localized FRBs with measured redshifts in the FRB host database\footnote{\url{http://frbhosts.org}}. Our sample thus consists of 7 repeating and 12 non-repeating FRBs. The data were primarily taken from the FRBSTATs catalog \citep{FRBSTAT}\footnote{\url{https://www.herta-experiment.org/frbstats/
}} and cross checked with original research papers. Details of the data sources are as follows: FRB 20180301A \citep{Price2019}; FRB 20180916B \citep{CHIME2021cat,Chawla2020,Marcote2020,Pilia2020}; FRB 20180924B \citep{Bannister2019}; FRB 20181030A \citep{CHIME2019b}; FRB 20181112A \citep{Prochaska2019}; FRB 20190102C, 20190608B, 20190611B, 20190711A \citep{Macquart2020};  FRB 20190614D \citep{Law2020}; FRB 20191228A, FRB 20200906A \citep{Bhandari2022}; FRB 20200120E \citep{Bhardwaj2021,Majid2021,Kirsten2022}; FRB 20201124A \citep{Farah2021, Herrmann2021, Hilmarsson2021, Marthi2021, KumarPra2022, Lanman2022}.  We have included all bursts with reliable $(T, F_\nu, \mathcal{F}_\nu)$ measurements for these 14 localized events \footnote{Four localized one-off FRBs 20190523A \citep{Ravi2019}, 20190714A \citep{Bhandari2019}, 20191001A \citep{Shannon2019} and 20200430A \citep{KumarPra2020b} are not adopted in the sample due to lacking in $T$ or $F_\nu$ value.}, where $\mathcal{F}_\nu$ is the burst fluence. In accordance with Paper I, we used the same data for FRB 20121102A \citep{LiD2021,Hessels2019,Cruces2021}. The FRB distance was obtained from the redshift information in the FRB host database, using the cosmological parameters $H_0=67.7\,\rm km\,s^{-1}\,Mpc^{-1}$, $\Omega_m=0.31$, $\Omega_\Lambda=0.69$ \citep{Planck2016}.

\begin{figure}
        \begin{center}
                \includegraphics[width=0.42\textwidth]{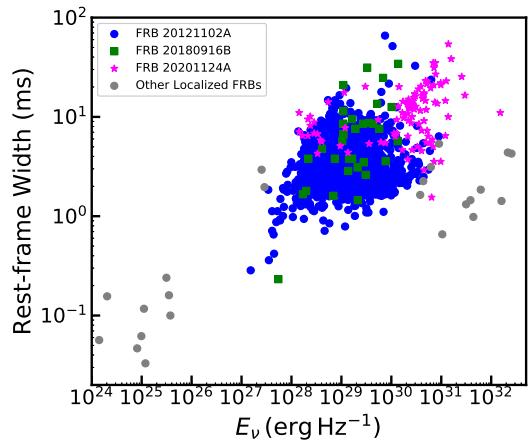}
                \caption{Distribution of bursts on the pulse width-energy plane for our full data sample. Blue, green, and magneta points represent three ``bursty'' repeating FRBs: 20121102A, 20180916B, and 20201124A respectively. Grey dots are for other localized FRBs. The overall distribution is quite scattered.}
                \label{fig1}
        \end{center}
\end{figure}

\begin{figure}
        \begin{center}
                \includegraphics[width=0.49\textwidth]{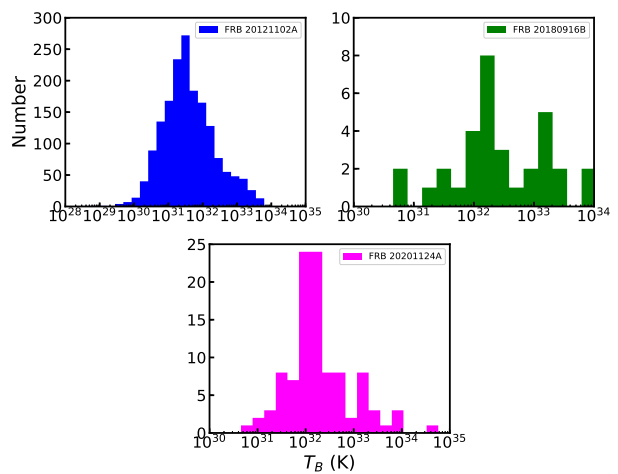}
                \caption{Number distribution of brightness temperatures for FRB 20121102A, 20180916B, and 20201124A in our sample.}
                \label{fig2}
        \end{center}
\end{figure}

Inspired by Paper I, we first investigate whether there is a direct correlation between observed pulse width and fluence for this large sample of bursts. However, since the redshifts of these FRB events are distinct, the cosmological expansion effect should be corrected and we obtain the rest-frame pulse width, $T_{\rm rest}$, after dividing $T$ by the $(1+z)$ factor. Furthermore, the fluence is an observed quantity, hence, here the radiated enengy should be relevant instead. We calculated the specific burst energy as $E_\nu=4\pi d_{\rm L}^2\mathcal{F}_\nu/(1+z)$. Typically, the total integrated energy can be estimated by multiplying $E_\nu$ with the observing bandwidth or the central frequency \citep{ZhangB2018,Aggarwal2021}, however, in our large sample, some bursts are in the extreme narrow band, while some others show continued emission beyond the observing bandwidth. We did not multiply either frequency in order to avoid extra bias. Figure \ref{fig1} shows the overall distribution of these FRBs on the $T_{\rm rest}-E_\nu$ plane. Obviously, this distribution is quite scattered, with no evidence of any correlation between the two quantities. Then we calculated the brightness temperature for each burst in the big sample. For three ``bursty'' FRBs, 20121102A, 20180916B, and 20201124A, we plotted their number distributions for $T_{\rm B}$ in Fig. \ref{fig2}. As in Paper I, we drew a dividing line by $T_{\rm B}$ for ``classical'' high-$T_{\rm B}$ FRBs. There is a sign that the brightness temperature distributions of FRBs 20180916B and 20201124A reach minimums around $10^{33}\,\rm K$, being close to the critical value for FRB 20121102A in Paper I. Therefore, as a trial, we adopted $T_{\rm B, cri}=10^{33}\,\rm K$ for all the FRB events in the sample. We note that the critical value is not necessarily the same for all FRBs, as can be seen in Table \ref{table1}.

We go on to investigate whether the pulse width-energy relation discovered in Paper I can be generalized to this large sample. After dropping the bursts with $T_{\rm B}<T_{\rm B,cri}$, we can plot the remaining bursts on the $T_{\rm rest}-E_\nu$ plane in Fig. \ref{fig3}. The blue, green, and magenta symbols represent ``bursty'' FRBs 20121102A, 20180916B, and 20201124A, respectively, while the grey dots are for other localized FRBs.  We can clearly see that $T_{\rm rest}$ is in positive relation with $E_\nu$, however, the slope varies for different events. We fit the correlation with simple power-laws and obtain the best-fitting lines for these three FRBs:
\begin{figure}
        \begin{center}
                \includegraphics[width=0.42\textwidth]{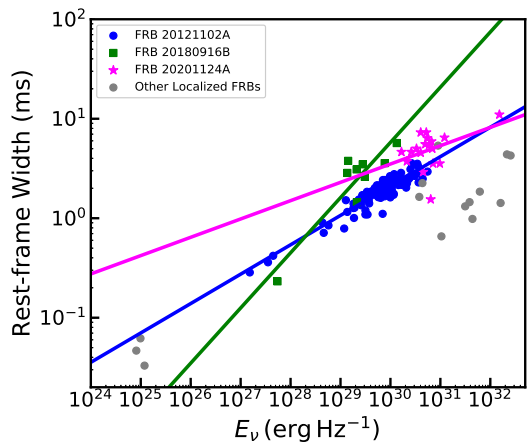}
                \caption{Distribution of bursts with $T_{\rm B}>10^{33}\,\rm K$ on the pulse width-energy plane. The positive power-law relations emerge for each FRB event, however, with different indexes. The reason is that $T_{\rm rest}$ is not really intrinsic and we need to remove the propagation effects to obtain the intrinsic width.}
                \label{fig3}
        \end{center}
\end{figure}

\bea
\log\left(\frac{T}{\rm ms}\right)&=& 0.30\log\left(\frac{E_\nu}{\rm erg\,Hz^{-1}}\right)-8.54, \quad\text{(20121102A),}\nonumber\\
\log\left(\frac{T}{\rm ms}\right)&=& 0.56\log\left(\frac{E_\nu}{\rm erg\,Hz^{-1}}\right)-15.88, \quad\text{(20180916B),}\nonumber\\
\log\left(\frac{T}{\rm ms}\right)&=& 0.18\log\left(\frac{E_\nu}{\rm erg\,Hz^{-1}}\right)-4.98, \quad\text{(20201124A),}
\label{eq:plfit}
\ena
respectively. These power-law relations are similar to that discovered with FAST sample alone in Paper I.

Furthermore, we sought to establish a universal relation for all FRBs. If this relation exists, it should be totally intrinsic and we need to remove all the propagation effects. As we know, the observed pulse width is broadened and should be a combined result \citep{Petroff2019},
\bea
T=\sqrt{T_{\rm i}^2(1+z)^2+t_{\rm samp}^2+\Delta t_{\rm DM}^2+\tau^2},
\label{eq:Tobs}
\ena
where the four terms on the right represent intrinsic width, data sampling time, dispersion smearing, and scattering time, respectively. Also, $t_{\rm samp}$ can be found for each survey. Dispersion smearing can be important if bursts are incoherently dispersed. For coherently dispersed bursts, the smearing timescale is usually tens of microseconds and can be neglected. We carefully picked out the incoherently-dispersed bursts for different surveys and calculated the relevant smearing. The scattering timescale $\tau$ is the most complicated step. For a large sample of different FRB events, the intervening medium between source and observer could make a huge difference, therefore, their scattering should be quite varied. Moreover, the motion of scattering medium is highly uncertain, leading to the observational fact that some bursts show clear scattering tails in their pulse profiles, while many others do not. 

In principle, the scattering of a burst is determined by Monte-Carlo fitting its pulse profile with an assumption of Gaussian intrinsic shape \citep{Ravi2019b, Qiu2020}. However, this is too time-consuming and almost unachievable for our large sample of more than two thousand bursts. Instead, we deal with scattering in a simple way below. The scattering is scaled with the frequency as a power law of $\tau\propto \nu^{\alpha}$, where $\alpha$ depends on the property of the scattering medium \citep{Lohmer2001,Cordes2003,Xu2016}. Furthermore, we assumed that the scattering time for  bursts in a same survey does not vary greatly. For instance, we assumed a single scattering time, $\tau_{\rm FAST}$, for the FAST sample of FRB 20121102A at the observing central frequency of 1.25 GHz. For other bursts of this event observed by Arecibo, GBT, and Effelsberg, the observing frequencies are different and the corresponding scattering can be determined by the frequency dependence. For FRBs 20180916B and 20201124A, we took the scattering of CHIME and GMRT observations as reference values, respectively. To avoid introducing too many free paremeters, we only adopted these three events here. Therefore, we introduced four parameters ($\tau_{\rm FAST},\,\tau_{\rm CHIME},\,\tau_{\rm GMRT}$, $\alpha$) to deal with scattering for our sample of bursts.

If a universal power-law relation between intrinsic pulse width and energy exists, we can parameterize it as:
\bea
\log T_{\rm i}=A\log E_\nu+B,
\label{eq:Tin}
\ena
where $A$ and  $B$ are two free parameters that are yet to be determined. Furthermore, as we point out above, the critical brightness temperature for classifying FRBs can vary for different FRB events, hence, we have three critical $T_{\rm B}$ values for these three FRBs. Therefore,  we have a total of nine free parameters, with which the observed pulse width can be expressed as:
\bea
T(E_\nu)=T(E_\nu;A,B,\tau_{\rm FAST},\tau_{\rm CHIME},\tau_{\rm GMRT},
T_{\rm B,20121102A}, \nonumber\\T_{\rm B,20180916B},T_{\rm B,20201124A},\alpha)
\label{eq:paras}
.\ena

\begin{figure}
\centering\includegraphics[angle=0,width=\linewidth]{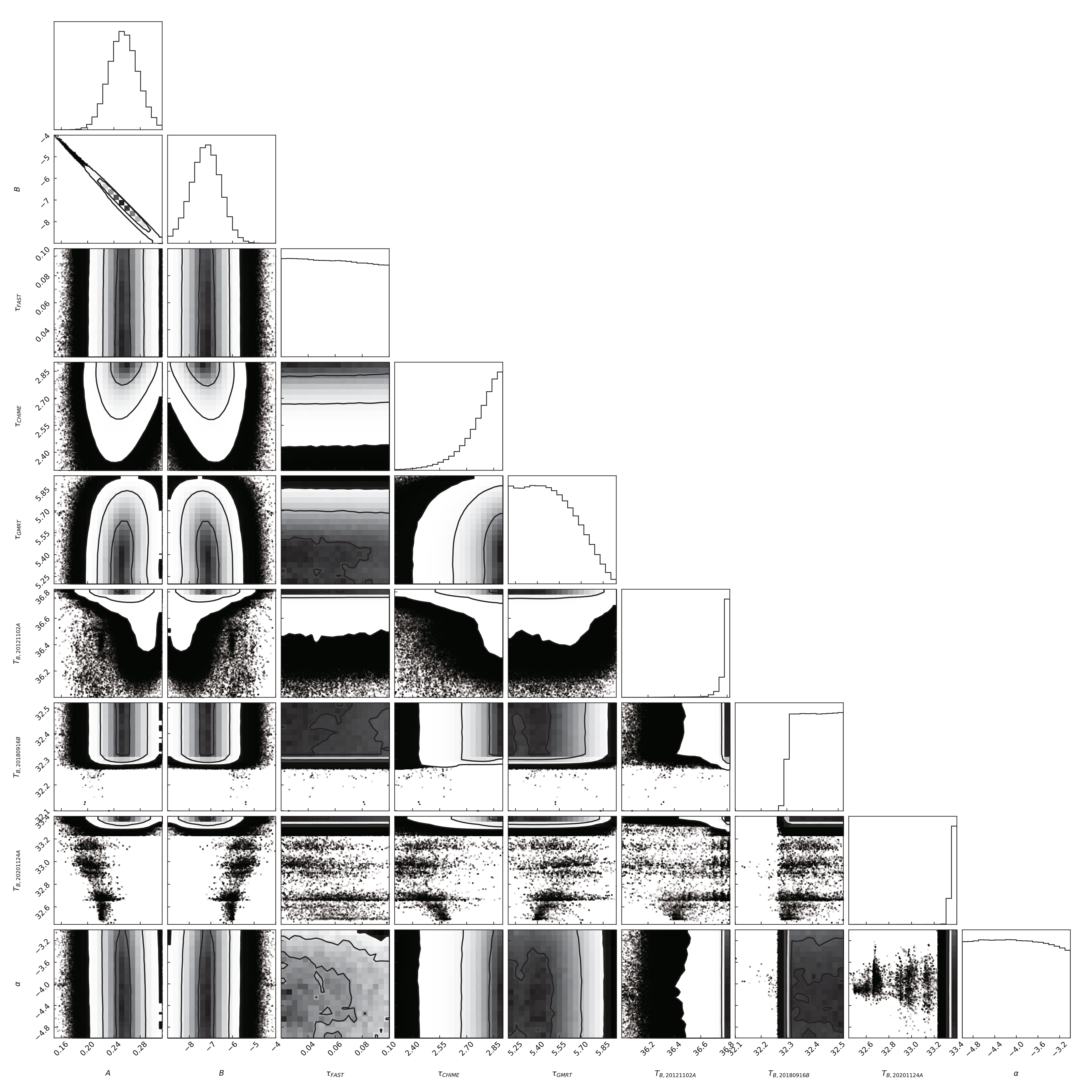} \ \
\caption{Parameter constraints on the nine parameters in Eq. \ref{eq:paras} using the emcee package. Histograms and contours illustrate the likelihood map.}
\label{fig4}
\end{figure}

\begin{figure}
        \begin{center}
                \includegraphics[width=0.42\textwidth]{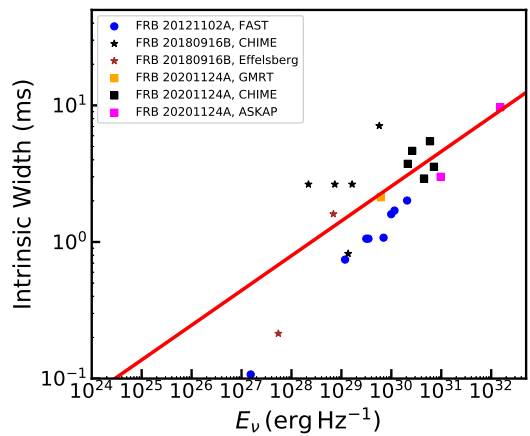}
                \caption{Relation between intrinsic pulse width and energy using an MCMC method. The best-fit critical lines $T_{\rm B,cri}$ for the three FRBs are different. After the classification, the remaining high-$T_{\rm B}$ bursts show a positive correlation of $T_{\rm i}\propto E_\nu^{0.25}$.}
                \label{fig5}
        \end{center}
\end{figure}

Next we carried out a Markov-Chain Monte-Carlo (MCMC) fitting of the observed pulse width using the emcee package \citep{Foreman-Mackey2013}. The prior (i.e., the allowed ranges in Table \ref{table1}) is set to a log uniform. We did not consider the intrinsic dispersion mainly because most of the bursts in the big sample are lacking in trustable error bars. Therefore, we defined a likelihood function that returns the sum of the squared difference. In this sense, the MCMC fitting result is nearly equivalent to that of the linear least-squares method. The results are shown in Table \ref{table1} and Fig. \ref{fig4} gives the relevant corner plot. We note that the best-fit values of three critical brightness temperatures always approach their higher ends, since this will eliminate more data points and the correlation is expected to be tighter with fewer bursts. To ensure there are enough data points left (at least five) for each repeater, we set the upper bounds for these three $T_{\rm B}$ artificially; this leads to low uncertainties. Using these best-fitting parameters, we can obtain the intrinsic width of bursts for the three FRB events. We plotted the classified bursts on $T_{\rm i}-E_\nu$ plane in Fig. \ref{fig5} and different symbols represent different events, with the discovery telescope also being marked. The red line corresponds to Eq.(\ref{eq:Tin}) with $A$ and $B$ given in Table \ref{table1}. Obviously there is evidence to assume a universal relation of $T_{\rm i}\propto E_\nu^{0.25}$ for these repeating FRBs.

\begin{table}
\centering
\caption{Best-fit values for the nine parameters.}
\label{table1}
\begin{tabular}{ccc}
\toprule Parameter &Allowed range & Best-fitting value \\
\midrule$A$ &[$0.0,\,1.0$]& $0.25_{-0.021}^{+0.022}$ \\
$B$ &[-10.0, 0.0] & $-7.22_{-0.69}^{+0.66}$ \\
$\tau_{\rm FAST}$ &[0.01, 0.1] & $0.059_{-0.027}^{+0.027}$ \\
$\tau_{\rm CHIME}$ &[2.0, 3.0]& $2.81_{-0.13}^{+0.067}$ \\
$\tau_{\rm GMRT}$ &[$5.0,\,6.0$]& $5.47_{-0.18}^{+0.21}$ \\
$\log T_{\rm B,20121102A}$ &[$32.0,\,36.82$]& $36.80_{-0.045}^{+0.011}$\\
$\log T_{\rm B,20180916B}$ &[$32.0,\,32.52$]& $32.41_{-0.076}^{+0.076}$\\
$\log T_{\rm B,20201124A}$ &[$32.0,\,33.40$]& $33.37_{-0.024}^{+0.019}$\\
$\alpha$ &[$-5.0,\,0.0$]& $-4.02_{-0.66}^{+0.68}$\\
\bottomrule & &
\end{tabular}%
\end{table}

The best-fit values in Table \ref{table1} are physically meaningful. The power-law index $A$ is close to the value we found using only FAST sample of FRB 20121102A \citep{XiaoD2022a}. By adopting the intrinsic width in this work, the positive correlation between $T$ and $E_\nu$ can be verified. The critical lines $T_{\rm B,cri}$ for FRB 20180916B and 20201124A are close to the nominal value of $\sim10^{33}\,\rm K$, implying that the sources for these two FRBs could be normal magnetars with a typical surface magnetic field $B_{\rm s}\sim10^{15}\,\rm G$ and rotational period of $P\sim 1\,\rm s$ \citep[see discussions in][]{XiaoD2022a}. However, the critical line for FRB 20121102A is nearly four orders of magnitude higher, corresponding to that the number of electrons in a single bunch is two orders of magnitude more abundant. Therefore the source of this FRB might be a more energetic magnetar for which the combination of $B_{\rm s}P^{-1}$ is greater by a factor of $\sim100$ \citep[see Eqs.8,9 of][]{XiaoD2022a}. The scattering power-law index is quite close to the expected values of $-4$ (for a thin extended scattering screen) and $-4.4$ (for a Kolmogorov spectrum of scattering medium) \citep{Xu2016}. This implies that the majority of bursts may be scattered with a scattering timescale that is only dependent on frequency.

\section{Consequences of the established correlation}
\label{sec3}
As we pointed out in Paper I, we can classify FRBs based on brightness temperature because there is an upper limit on $T_{\rm B}$ for each radiation mechanism. Generally, the critical lines for coherent curvature radiation and synchrotron maser radiation lie around $T_{\rm B,cri}\sim10^{33}\,\rm K$ \citep{XiaoD2022a}, which is consistent with the best-fit critical $T_{\rm B}$ values of FRB 20180916B and 20201124A (shown in Table \ref{table1}). Below $T_{\rm B,cri}$, the correlation of $T_{\rm i}\propto E_\nu^{0.25}$ is buried since multiple mechanisms can produce those bursts. Beyond $T_{\rm B,cri}$, we can expect one single mechanism to be at work, therefore, the above correlation emerges.

This correlation certainly gives us some hints on the FRB radiation mechanism and we can explore this issue in a preliminary fashion:\ basically, for the coherent curvature radiation, the duration of emission is in proportion to the single bunch length, $l_\parallel$, and the number of bunches $N_{\rm b}$ \citep{Kumar2017}:
\bea
T_{\rm i}\sim N_{\rm b}\frac{\lambda}{c}\propto N_{\rm b}\nu^{-1},
\label{eq:duration}
\ena
where the bunch length, $l_\parallel$, is on the same order as the emission wavelength $\lambda$. Meanwhile, the specific energy can be approximated as
\bea
E_\nu\sim L_{\rm curv}T_{\rm i}/\nu,
\label{eq:fluence}
\ena
where the luminosity of coherent curvature radiation is
\bea
L_{\rm curv}\sim N_{\rm b}N_{\rm e}^2\gamma^2 P_{\rm curv},
\label{eq:Lcurv}
\ena
assuming $N_{\rm e}$ electrons moving with Lorentz factor $\gamma$ in each bunch. Substituting the single electron emission power $P_{\rm curv}=\frac{2}{3}\frac{\gamma^4e^2c}{\rho^2}$ and replacing the curvature radius $\rho$ with the characteristic frequency, $\nu=\frac{3c\gamma^3}{4\pi\rho}$, we obtain:
\bea
E_\nu\propto N_{\rm b}N_{\rm e}^2\gamma^2\frac{\gamma^4}{(\gamma^3/\nu)^2}N_{\rm b}\nu^{-1}/\nu\propto N_{\rm b}^2N_{\rm e}^2.
\ena
The number of electrons in the coherently emitting region is in proportion to its volume: $V_{\rm coh}=\pi l_\parallel \min[l_\perp^2,(\gamma\lambda)^2]$, where $l_\perp$ is the transverse size of the bunch \citep{Kumar2017}. Therefore, we have

\bea
E_\nu\propto
\begin{cases}
N_{\rm b}^2l_\perp^4\nu^{-2}, &l_\perp\leq\gamma\lambda,\\
N_{\rm b}^2\gamma^4\nu^{-6}, &l_\perp>\gamma\lambda.
\end{cases}\label{eq:curv_corr}
\ena
Assuming that $l_\perp$ does not vary substantially, then we can expect $T_{\rm i}\propto E_\nu^{0.5}$ for the case $l_\perp\leq\gamma\lambda$. Similarly, $T_{\rm i}\propto E_\nu^{1/6}$ is obtained for the other case if we assume both $N_{\rm b}$ and $\gamma$ have narrow value ranges (although somewhat rigorous).

The above two scalings deviate from the correlation we found, implying once again that coherent curvature radiation is unlikely to be the mechanism for high-$T_{\rm B}$ ``classical'' bursts. Currently, the FRB radiation mechanism is largely unknown and it was only recently that \citet{Zhang2022} proposed that coherent inverse Compton scattering (ICS) could produce very high-$T_{\rm B}$ bursts attributed to enhanced single electron emission power. Here, we examine the expected scaling for this mechanism in a similar way. The duration of emission is the same with Eq. \ref{eq:duration}. The luminosity of coherent ICS has the similar form to Eq. \ref{eq:Lcurv}, with $P_{\rm curv}$ substituted by the ICS power, $P_{\rm ICS}=\frac{4}{3}\gamma^2\sigma_{\rm ICS}cU_{\rm ph}$. The scattering cross-section is proportional to $\gamma^{-2}$ and the photon energy density does not vary greatly if we assume the scale of crust oscillation basically remain unchanged \citep{Zhang2022}. Therefore, for this mechanism, we have:
\bea
E_\nu\propto N_{\rm b}^2N_{\rm e}^2\nu^{-1}\propto\begin{cases}
N_{\rm b}^2l_\perp^4\nu^{-3}, &l_\perp\leq\gamma\lambda,\\
N_{\rm b}^2\nu^{-5}, &l_\perp>\gamma\lambda,
\end{cases}\label{eq:ICS_corr}
\ena
with the ICS emission frequency, $\nu\propto\gamma^2$, is substituted. If we assume $N_{\rm b}$ and $l_\perp$ remain unchanged, then we can expect $T_{\rm i}\propto E_\nu^{1/3}$ and $T_{\rm i}\propto E_\nu^{1/5}$  for the above two cases respectively.

In reality, the number of bunches and bunch size can not be all the same from burst to burst. However, as long as they have relative small value ranges, the positive correlation between $T_{\rm i}$ and $E_\nu$ is still expected for both the coherent curvature radiation and ICS radiation mechanisms. The power-law index of $\sim$0.25 that we found lies well between $1/5$ and $1/3$, which might be ascribed to the variation of $N_{\rm b}$. Therefore, we consider coherent ICS radiation as a probable mechanism for inducing high-$T_{\rm B}$ ``classical'' FRBs.

The scaling of $T_{\rm i}$ with $E_\nu$ is not straight-forward for synchrotron maser radiation. In the scenario described by \citet{Metzger2019}, the maser emission has been reprocessed by the external medium, then the observed peak frequency is higher than the intrinsic maser frequency. Therefore, the observed FRB energy strongly depends on the density of the ambient medium \citep{XiaoD2020}. The calculated burst energy does not vary monotonously with the medium density \citep[see Fig. 6 in][]{Metzger2019}. Also, the intrinsic maser spectrum strongly depends on the upstream magnetization and can be only given by detailed particle-in-cell simulations \citep{Plotnikov2019}. It is for these reasons that we do not expect a simple relation between width and energy for this mechanism. A more careful treatment is needed to find out whether there exists any correlation between other physical quantities for synchrotron maser mechanism, as well as for other proposed radiation mechanisms \citep{Waxman2017,Wadiasingh2019,Lyubarsky2020,Lyutikov2021a}

\section{Discussion and conclusions}
\label{sec4}
In this paper, we present a detail analysis of the burst width-energy correlation using a large sample of FRBs. We found a positive relation of $T_{\rm i}\propto E_\nu^{0.25}$ for three bursty FRBs, confirming the results given in Paper I, based on a single repeating FRB. We note that we used the observed width from the previous paper. In a physical sense, this is not very accurate since our classification criterion $T_{\rm B}$ is an intrinsic property that relates directly to the radiation mechanism. In order to look for any physical correlation, we need to remove all the propagation effects and get the intrinsic width. However, the relation between the observed width and fluence in Paper I is still tenable if the scattering time between individual bursts does not vary greatly at a given observation frequency for FRB 20121102A.

Based on the claim that brightness temperature is a more feasible classification criterion for FRBs than\ repetition since it is determined by the radiation mechanism, we investigated two mainstream mechanisms. We found that both coherent curvature radiation and synchrotron maser radiation cannot easily reach high brightness temperature. In addition, we explain the $T_{\rm i}-E_\nu$ relation, finding that coherent ICS by bunches can meet these requirements, but still remain to be verified by further observational evidence.

There may be several factors that can influence the power-law index of this correlation. First, the method of obtaining intrinsic pulse width has been handled in a simple way. We assumed that all bursts are scattered, with their scattering time dependent only on the frequency in a power-law form. In fact, many bursts do not show scattering tails in their burst profiles, therefore, the scattering time should be fit from burst to burst in order to be accurate. Second, selection effects and observational bias can play a role. The measured width and fluence are only lower limits if FRB radiation spectrum partially lies in the observing band. Observationally, many bursts show a trend of continued emission beyond the observing band in their waterfall plots, while statistical analyses of narrow-banded bursts fully within the observing band may approach the realistic scenario more closely \citep{Aggarwal2021}. In addition, a ``tip-of-iceberg'' effect might also be at work for weak bursts. The measured pulse width could be shorter than the intrinsic value for a weak burst as its flux falls below the background noise. In this sense, bursts with high signal-to-noise ratios are preferred.

Except for the $T_{i}-E_\nu$ relation, other empirical two-parameter correlations of classified FRBs can be searched for in future studies, in a similar to those of supernovae and gamma-ray bursts \citep[e.g.,][]{Amati2002}. We note that the luminosity-duration relation has already been discussed \citep{Hashimoto2019,Hashimoto2020}. These correlations make FRBs potential ``standard candles'' that can be very useful in cosmological studies, as long as the scatter of the correlation can be effectively reduced when using a greater store of data in the future.

\begin{acknowledgements}
This work is supported by the National Key Research and Development Program of China (Grant No. 2017YFA0402600), the National SKA Program of China (grant No. 2020SKA0120300), and the National Natural Science Foundation of China (Grant No. 11833003, 11903018). DX is also supported by the Natural Science Foundation for the Youth of Jiangsu Province (Grant NO. BK20180324) and Shanghai Sailing Program (Grant No. 19YF1420300).
\end{acknowledgements}

\bibliographystyle{aa} % style aa.bst
\bibliography{FRBlatest} % your references Yourfile.bib

\end{document}